# Optical and transport properties of short period InAs/GaAs superlattices near quantum dot formation

V A Kulbachinskii†, R A Lunin†, V A Rogozin†, V G Mokerov‡,

Yu V Fedorov‡, Yu V Khabarov‡ and A de Visser§

†Low Temperature Physics Department, Moscow State University,

119899, Moscow, Russia

‡Institute of Radioengineering and Electronics, Russian Academy of Sciences,

103907 Moscow, Russia

§Van der Waals-Zeeman Institute, University of Amsterdam,

Valckenierstraat 65, 1018 XE Amsterdam, The Netherlands

**Abstract.** We have investigated the optical and transport properties of MBE grown short-period superlattices of InAs/GaAs with different numbers of periods ($3 \leq N \leq 24$) and a total thickness 14 nm. Bandstructure calculations show that these superlattices represent a quantum well with average composition $In_{0.16}Ga_{0.84}As$. The electron wave functions are only slightly modulated by the superlattice potential as compared to a single quantum well with the same composition, which was grown as a reference sample. The photoluminescence, the resistance, the Shubnikov-de Haas effect and the Hall effect have been measured as a function of the InAs layer thickness $Q$ in the range $0.33 \leq Q \leq 2.7$ monolayers (ML). The electron densities range from 6.8 to $11.5 \times 10^{11}$ cm$^{-2}$ for $Q \leq 2.0$ ML. The photoluminescence and magnetotransport data show that only one subband is occupied. When $Q \geq 2.7$ ML quantum dots are formed and the metallic type of conductivity changes to variable range hopping conductivity.

Corresponding author:
  Dr. A. de Visser
  Van der Waals-Zeeman Institute, University of Amsterdam,
  Valckenierstraat 65, 1018 XE Amsterdam, The Netherlands
  Phone: +31-20-5255732/5663; Fax: +31-20-5255788
  E-mail: devisser@science.uva.nl

## 1. Introduction

In recent years much research has been focussed on self-organized ensembles of quantum dots. Such quantum dots are quasi-zero-dimensional objects (with sizes ~ 5-20 nm), which form during hetero-epitaxial growth due to the mismatch of the lattice parameters of the grown material and the substrate. The mechanism of nucleation and formation of quantum dots in the Stransky-Krastanov growth mode has been studied intensively [1-6]. The process of self-organized growth of InAs quantum dots on the surface of GaAs starts when the thickness $Q$ of the InAs layer exceeds some critical value. There is a large amount of literature dealing with the investigation of the optical properties of undoped structures in the quantum dot regime [3,6]. However, structures with InAs layer thicknesses below the threshold for quantum dot formation are less studied. Here we report on the optical and electrical properties of *doped* structures at the initial stage of quantum dot formation. Short period InAs/GaAs superlattices with periods $3 \leq N \leq 24$ and total thickness 14 nm were prepared with very thin InAs layers ranging from 0.33 to 2.7 monolayers (ML). Energy band calculations show that the wave functions of the superlattice are equivalent to those of a single quantum well, with a small modulation due to the periodic potential of the superlattice. In order to investigate the optical and electrical properties of these superlattices we measured the photoluminescence at a temperature $T = 77$ K and the resistance $R$ in the $T$-interval 1-300 K. In addition, Shubnikov-de Haas (SdH) and Hall effect measurements were carried out in a long-pulse (1 s) high-field magnet at the University of Amsterdam. The maximum field was 40 T and the $T$-interval is 1.4-4.2 K.

## 2. Samples

The short-period superlattice structures were grown by molecular beam epitaxy (MBE) at a temperature of 490 °C on semi-insulating (001) GaAs substrates. A schematic drawing of the structure is given in Fig.1. The $N$ period InAs/GaAs superlattices, with a total thickness of 14 nm, were grown on a 1 µm thick GaAs buffer layer. The value 14 nm of the total thickness is dictated by the optimal composition $In_{0.16}Ga_{0.84}As$ from the point of view of the quality of the crystal structure. Larger thicknesses in strained structures with the same composition may give rise to lattice defects. The effective thickness $Q$ of the InAs layers ranged from 0.33 to 2.7 ML, while the thickness $d$ of the GaAs layers in the superlattice was varied correspondingly from 1.7 to 13.5 ML, as to keep the mean composition of the superlattice equivalent to that of the solid solution $In_{0.16}Ga_{0.84}As$. The Si δ-doping layer, with a Si sheet concentration $5\times10^{12}$ cm$^{-2}$, was separated from the superlattice by a 10 nm thick spacer layer of composition $Al_{0.2}Ga_{0.8}As$. Subsequently, a 35 nm thick layer of $Al_{0.2}Ga_{0.8}As$ was grown, followed by a 6 nm thick GaAs cap layer. As reference sample, a single quantum well $In_{0.16}Ga_{0.84}As$ (well width 14 nm) was grown.

In order to measure in-plane electron transport, Hall bars were prepared from the wafers for all structures. The (double) Hall bars were L-shaped as to allow the current $I$ to flow parallel (∥) and perpendicular (⊥) to the [110] direction in the same sample. This allows for an accurate investigation of anisotropy in the transport properties. The relevant parameters of the structures are listed in Table 1. Photoluminescence and atomic force microscopy (AFM) data clearly demonstrate that for $Q < 2.7$ ML nanoscale semiconductor islands of InAs are formed, while for $Q \geq 2.7$ ML InAs quantum dots form. The latter is illustrated in Fig.2, which shows the AFM image of sample #8 ($Q = 2.7$ ML) after selective etching of the surface.

## 3. Results and discussion

*3.1 Temperature dependence of the resistivity*

The temperature dependence of the sheet resistivity $R_o(T)$ of some selected structures is shown in Fig.3. For the samples with $Q \leq 2.0$ ML the resistivity has a very weak temperature dependence with an absolute value of the sheet conductivity more than $e^2/h$, i.e. the critical conductivity value for the transition from the 2D metallic phase to the insulating phase. At low temperatures the conductivity shows a logarithmic temperature dependence, $s_o \sim \ln T$, which is characteristic of weak localization [7]. The ln$T$ behaviour is typically valid in the range $2\,K < T < 15\,K$, as is illustrated for samples #2 and #4 in Fig.4. At lower $T$ the conductivity starts to saturate. The reason for this saturation is not clear.

The resistivity of the structure with $Q = 2.7$ ML (sample #8) behaves very differently. $R_o(T) \sim h/e^2$ at room temperature and increases with decreasing temperature. This is in line with the formation of quantum dots in the layers. At the lowest temperatures the resistivity enters the variable range hopping conductivity regime for the 2D case, $R_o(T) \sim \exp[(T_0/T)^{1/3}]$ [8], as is shown in the insert in Fig. 4.

*3.2 Photoluminescence*

In Fig.5 we show the photoluminescence (PL) spectra for the different structures. For samples with $Q \leq 2.0$ ML two peaks are observed: i) a low-energy peak with a maximum at photon energies 1.356-1.375 eV, and ii) a high-energy peak with a maximum at photon energies 1.406-1.434 eV (see also Table 1). These two peaks correspond to optical transitions from the two electronic subbands, i.e. from the occupied lower subband and the unoccupied upper subband (see section 3.4), to the hole subband [9]. The intensity of the low-energy peak is always higher. This is similar to the case of undoped quantum wells, where the optical transition from the lowest subband is readily observed in PL spectra, while the transition from the upper subband is very weak. A distinctly different behaviour is observed when the effective thickness of the InAs layers $Q \geq 2.7$ ML. A new broad and intense photoluminescence peak with a maximum at 1.265 eV appears in the long-wavelength region of the photoluminescence spectrum (see data for sample #8 in Fig.5). As shown in Ref.[1-3], such a transformation of the photoluminescence spectrum can be attributed to the change from 2D layer-by-layer growth to the formation of vertically-stacked quantum dots. The presence of quantum dots in sample #8 was confirmed by AFM (see Fig. 2).

A second noteworthy feature of the photoluminescence spectra is the non-monotonous dependence of the intensity $I_{PL}$ on the InAs layer thickness $Q$. The largest intensities are observed for sample #2 ($Q = 0.33$ ML) and #7 ($Q = 2.0$ ML). For the same samples we observe maxima of the electron mobility $m$ (see Table 1).

*3.3 The energy spectrum*

For all structures, the electron wave functions $\psi_i(z)$ and the energies $E_i$ were determined in the effective mass approximation by the self-consistent solution of the one-dimensional Schrödinger and Poisson equations [10,11]. In the Schrödinger equation

$$\left[ -\frac{\hbar^2}{2}\frac{d}{dz}\left(\frac{1}{m^*(z)}\frac{d}{dz}\right) + U(z) \right] \psi_i(z) = E_i \psi_i(z) \tag{1}$$

the potential energy $U(z) = U_c(z) + U_H(z) + U_{xc}(z)$ is the sum of the conduction band discontinuity $U_c(z)$ at the heterojunctions, the electrostatic potential energy $U_H(z)$ (Hartree potential) and the exchange-correlation potential $U_{xc}(z)$. The electrostatic potential energy is determined by the Poisson equation

$$\frac{d}{dz}\left(\varepsilon_0 \varepsilon(z)\frac{dU_H(z)}{dz}\right) = e^2 [N(z) - n(z)] \tag{2}$$

where $\varepsilon_0$ is the permittivity of the vacuum, $\varepsilon$ is the permittivity of the material, $N(z)$ is the 3D density of ionized donors and

$$n(z) = \frac{m^*}{\pi\hbar^2}\sum_i [E_F - E_i]\,\theta[E_F - E_i]\,|\psi_i(z)|^2 \tag{3}$$

is the electron concentration at $T = 0$ K, with $\theta(x)$ the Heaviside function. The exchange-correlation potential was approximated by the formula [12]

$$U_{xc} = -\left[1 + 0.0545\, r_S \ln\left(1 + \frac{11.4}{r_S}\right)\right]\frac{2}{\pi \alpha r_S} Ry^* \tag{4}$$

where

$$\alpha = \left(\frac{4}{9\pi}\right)^{1/3}, \quad r_S = \left(\frac{4\pi a_B^{*3} n(z)}{3}\right)^{-1/3}, \quad a_B^* = \frac{4\pi\varepsilon_0 \varepsilon \hbar^2}{m^* e^2}, \quad Ry^* = \frac{e^2}{8\pi\varepsilon_0 \varepsilon a_B^*} \tag{5}$$

The values of the conduction and valence bands discontinuities of GaAs and strained InAs were taken equal to $\Delta U_c = 0.535$ eV and $\Delta U_v = 0.385$ eV [13], respectively. The electron effective mass in strained InAs is equal to $m_e^{st} = 0.0365\, m_0$ [14].

Using the calculated profile of the conduction band, we determined the profile of the valence band and the hole energy levels in the quantum well. Typical examples of the profile of the bottom of the conduction band $E_c$, the position of the two subbands $E_0$ and $E_1$ and the corresponding wave functions are shown in Fig.6 for samples #4 and #7. The energy is measured with respect to the Fermi level $E_F$. In all samples only the lowest energy level is occupied by electrons. This is confirmed by the SdH data (see section 3.4). The calculated energies of the optical transitions are in good agreement with the positions of the peaks in the photoluminescence spectra. Fig. 6 shows that the wave functions of the superlattices are almost identical to those of a single quantum well with the average composition $In_{0.16}Ga_{0.84}As$ and the same well width. The difference is a small modulation due to the superlattice potential. We conclude that very short period superlattices behave like a single quantum well with a small modulation of the potential.

*3.4 Magnetoresistance*

For all samples we observed a negative magnetoresistance at low temperatures in low magnetic fields ($B < 0.1$ T). Such a dependence is expected for the weak localization regime in the 2D case [7], which was inferred from the logarithmic decrease of the conductivity with decreasing temperature (see Fig.4).

In Fig.7 we show high-field transport data for samples #4 and #6. The longitudinal resistance shows SdH oscillations, while the quantum Hall effect is observed for fields exceeding ~7 T. The transversal resistance $R_{xy}$ shows clear plateau values at the integer filling fractions **n** = 1, 2 and 3. The Fast Fourier Transform (FFT) of the SdH signal in the low-magnetic field data ($B$ À 8 T, see insert in Fig.7) shows one single frequency, which implies that only the lowest subband is occupied. The carrier concentrations determined from the period of the SdH effect coincide with those determined from the low-field Hall effect and fall in the range $n = 1.5\text{-}12 \times 10^{11}$ cm$^{-2}$ for the different structures at $T = 4.2$ K.

The calculated mobilities **m** for the different structures with $Q \leq 2$ ML range from 0.21 to 0.94 m$^2$/Vs at $T = 4.2$ K. The highest mobility values, **m** = 0.94 m$^2$/Vs and **m** = 0.71 cm$^2$/Vs are observed for sample #2 and #7, respectively. These relatively high mobilities are possibly explained by the weakness of the elastic strain fluctuations and a near-perfect crystal lattice as compared to the solid solution In$_{0.16}$Ga$_{0.84}$As. It is noteworthy that for these two samples also the intensity of the photoluminescence peak is much higher. In sample #8 quantum dots are formed and the electron concentration is reduced to $n = 1.5 \times 10^{11}$ cm$^{-2}$, i.e. much smaller than in the other structures. Consequently, the electron mobility **m** = 0.005 m$^2$/Vs is very low.

*3.5 Anisotropy of conductivity*

For all superlattice samples the transport experiments on the L-shaped Hall bars signal a significant anisotropy of the resistivity. Anisotropy was not detected in the transport measurements on the quantum well with composition In$_{0.16}$Ga$_{0.84}$As (sample #1). The ratio of the resistance $R_\perp/R_\parallel$ depends on the thickness $Q$ and attains a values up to ~1.3. The corresponding anisotropic electron mobility is related to an asymmetric dislocation distribution [15]. The anisotropy of the resistance in 2D electron systems is typical for structures with preferential growth of deposited material in one direction (see e.g. Refs.16 and 17). The observed $Q$ dependence of the anisotropy indicates that elongated island growth sets in above a certain threshold value [18].

**4. Conclusions and summary**

The optical and transport properties of MBE grown short-period superlattices of InAs/GaAs with different numbers of periods ($3 \leq N \leq 24$) have been studied. The total thickness of the superlattice is 14 nm and was kept constant by varying the thickness $Q$ of the InAs layer and $d$ of the GaAs layer. The electronic wave functions were calculated self-consistently by solving the one-dimensional Schrödinger and Poisson equations. For $Q \leq 2.0$ ML our superlattices can be described as a single quantum well with average composition $In_{0.16}Ga_{0.84}As$, with electronic wave functions slightly modulated due to the superlattice potential. The photoluminescence and the high-field Shubnikov-de Haas effect show that only the lowest subband is occupied. For some of the structures we observe the quantum Hall effect. For $Q > 2.0$ ML the photoluminescence and the variable range hopping conductivity resistivity data provide further evidence for the formation of quantum dots. The photoluminescence peak and the electron density and mobility do not depend monotoneously on the InAs layer thickness. For $Q = 0.33$ ML and $Q = 2.0$ ML enhanced values for the intensity of the photoluminescence peak and mobility are observed. This is possibly due to a higher quality of these structures, due to a more effective relaxation of the strain. A small but significant anisotropy in the resistance was observed which depends on the thickness of the InAs layers, as expected for elongated island growth.

The work was supported by the RFBR under grants numbers 00-02-17493 and 01-02-17748.

**Table 1** Structural and electronic parameters of short-period InAs/GaAs superlattices with a total thickness 14 nm. $Q$ and $d$ are the InAs and the GaAs layer thicknesses expressed in monolayers (ML), respectively. $N$ is the number of superlattice periods. Maxima in the photoluminescence spectra are observed at h$n_{max}$ (at $T = 77$ K). $n$ is the carrier density determined by the Hall effect and $m$ is the mobility (at $T = 4.2$ K).

| Sample number | $Q$ (InAs, ML) | $d$ (GaAs, ML) | Number of periods $N$ | h$n_{max}$ (eV) | $n$ ($10^{11}$ cm$^{-2}$) | $m$ (cm$^2$/Vs) |
|---|---|---|---|---|---|---|
| 1 | Quantum well In$_{0.16}$Ga$_{0.84}$As | Quantum well In$_{0.16}$Ga$_{0.84}$As | - | 1.434, 1.375 | 8.1 | 8100 |
| 2 | 0.33 | 1.7 | 24 | 1.419, 1.367 | 11.5 | 9400 |
| 3 | 0.67 | 3.4 | 12 | 1.411, 1.369 | 7.2 | 2060 |
| 4 | 1.00 | 5.0 | 8 | 1.411, 1.370 | 7.3 | 2450 |
| 5 | 1.33 | 6.7 | 6 | 1.418, 1.374 | 8.7 | 4220 |
| 6 | 1.58 | 8.0 | 5 | 1.404, 1.368 | 6.8 | 4910 |
| 7 | 2.00 | 10.0 | 4 | 1.406, 1.356 | 10.4 | 7060 |
| 8 | 2.70 | 13.5 | 3 | 1.390, 1.265 | 1.5 | 50 |

**Figure captions**

Fig. 1 Schematic drawing of the short-period InAs/GaAs superlattice structures. The number of periods is $3 \leq N \leq 24$. The InAs layer has a thickness $Q$ and the GaAs layer has a thickness d. The total width of the superlattice is 14 nm.

Fig. 2 Atomic force microscopy image of sample #8 ($Q = 2.7$ ML) after selective etching of the cap layer. Quantum dots are formed.

Fig. 3 Temperature dependence of the sheet resistance for samples #1-#4, #6 and #8. The dashed line corresponds to the critical value of the resistivity $h/e^2$.

Fig.4 Temperature dependence of the sheet conductivity (in units of $e^2/h$) for samples #2 and #4. The data follow a $\ln T$ dependence between ~2 and 15 K. The insert shows the low-temperature conductivity for sample #8 (with quantum dots). The data follow the Mott law for variable range hopping conductivity: $R_o \sim \exp(T_0/T)^{1/3}$ (solid line).

Fig. 5 Photoluminescence spectra of short-period InAs/GaAs superlattices with InAs layer thickness 0.33 ML $\leq Q \leq$ 2.7 ML. For comparison the spectrum for a single quantum well with composition $In_{0.16}Ga_{0.84}As$ (sample #1) is shown as well. The arrows indicate the positions of maxima.

Fig. 6 Calculated band diagram for superlattice structures #4 with 4 periods (a) and #7 with 8 periods (b). $E_c$ denotes the bottom of the conduction band. The lowest two subbands $E_0$ (dashed line) and $E_1$ (dash-dotted line) are indicated together with the corresponding wave functions. The Fermi level is represented by the solid line.

Fig. 7 High-field magnetotransport data for samples #4 with $Q = 1.0$ ML (a) and #6 with $Q = 1.58$ ML (b) taken at $T = 4.2$ K. $R_{xxo}$ is the longitudinal sheet resistance and $R_{xy}$ the transversal resistance. The quantum Hall effect becomes visible above fields of ~8 T. The inserts show the Fast Fourier Transform of the SdH signal.

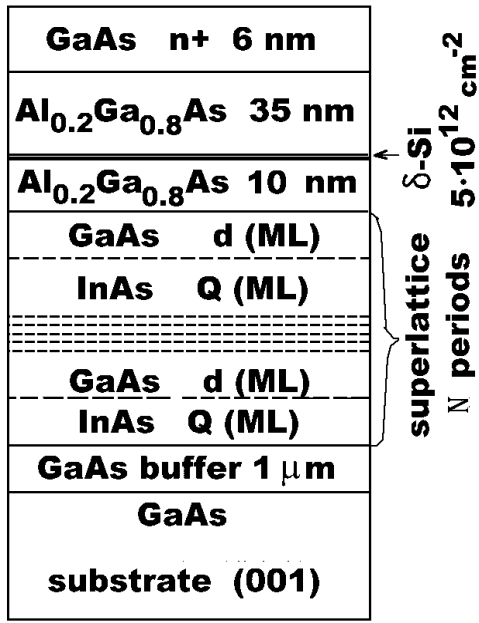

Fig. 1

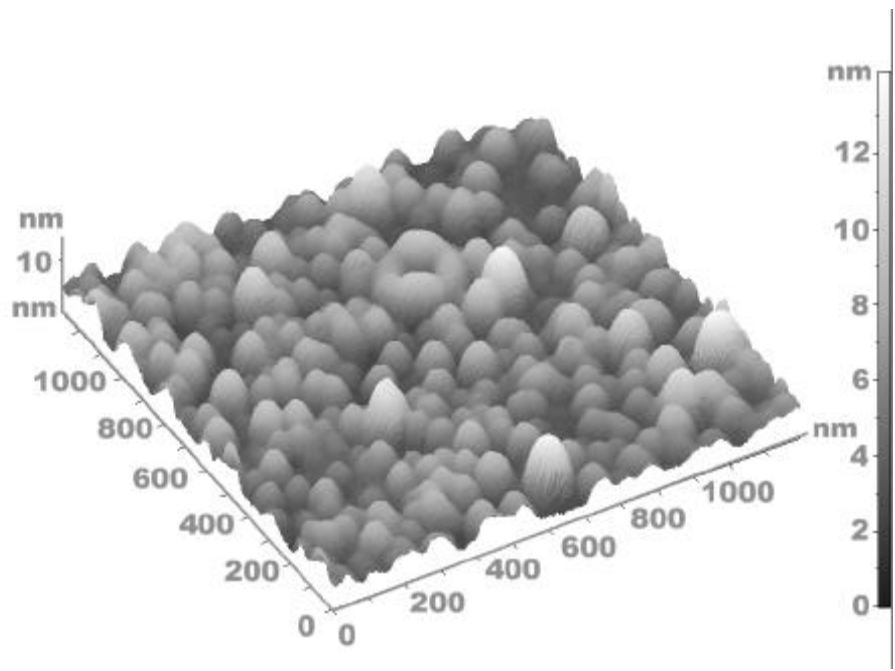

Fig. 2

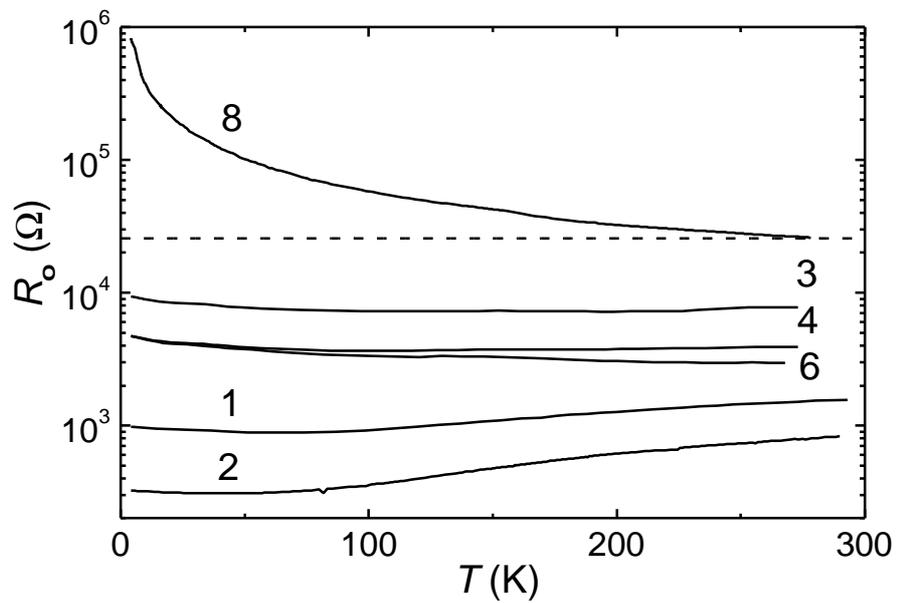

**Fig.3**

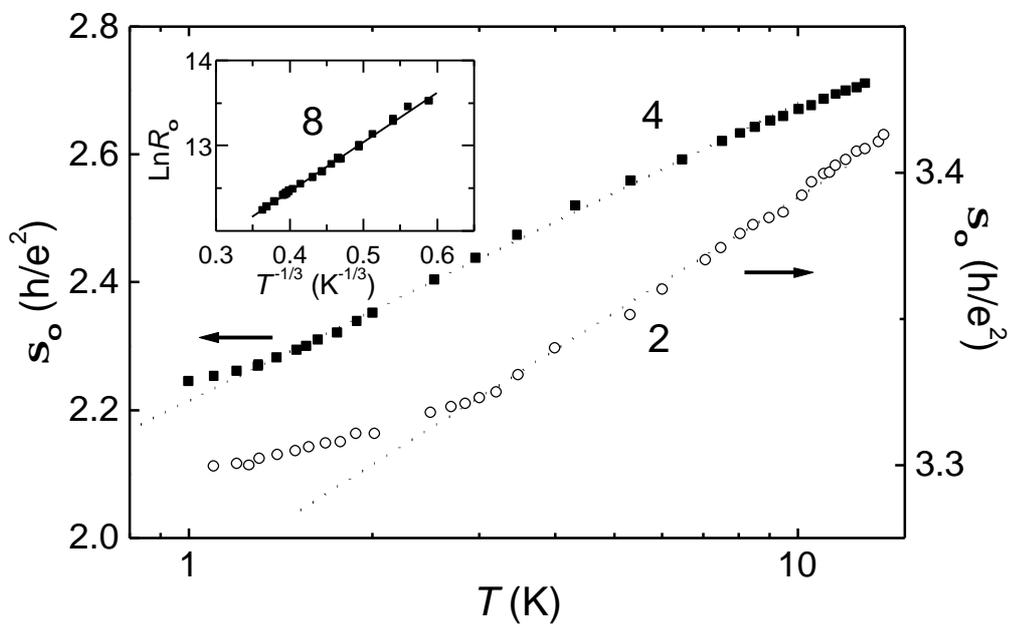

**Fig. 4**

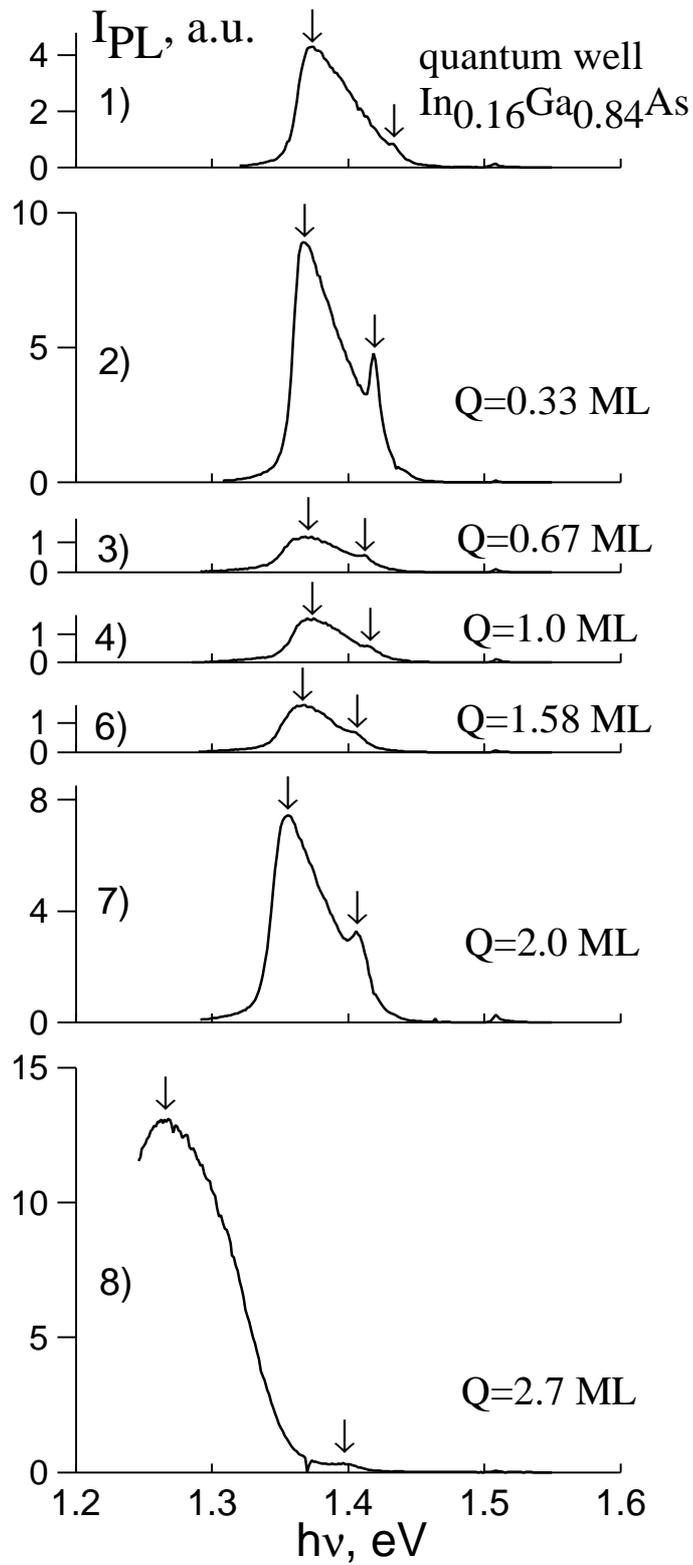

**Fig. 5**

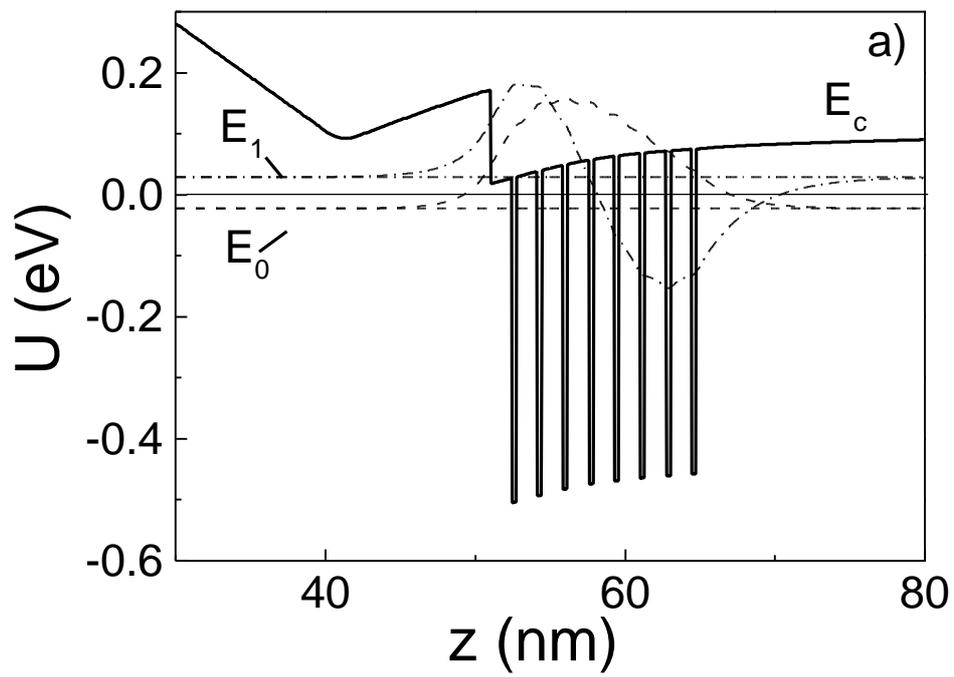
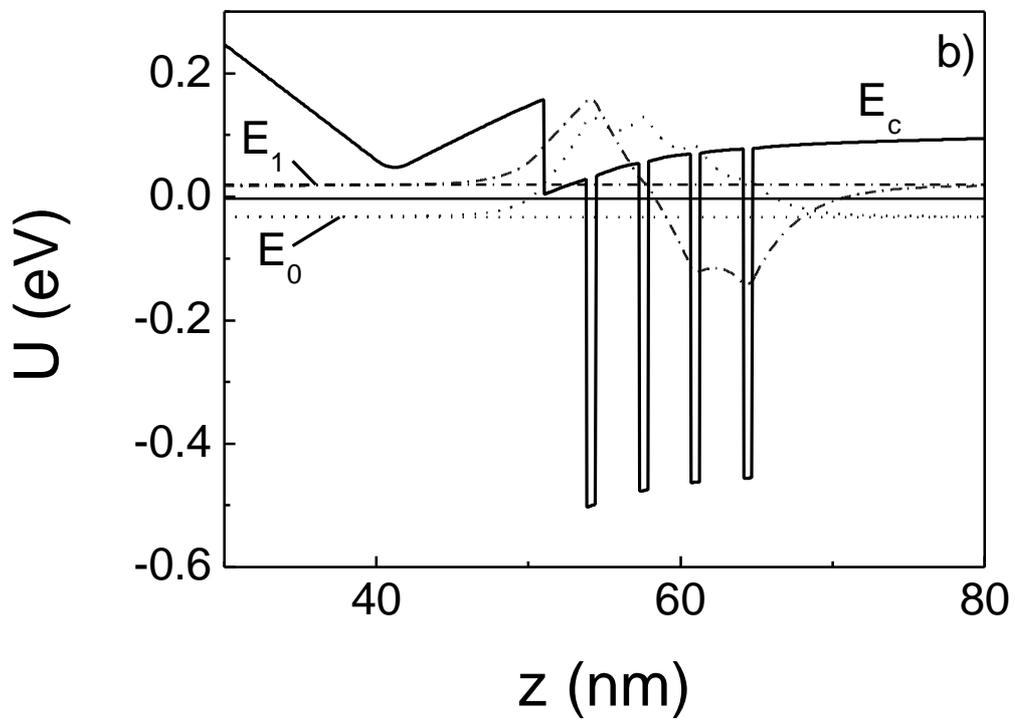

**Fig.6**

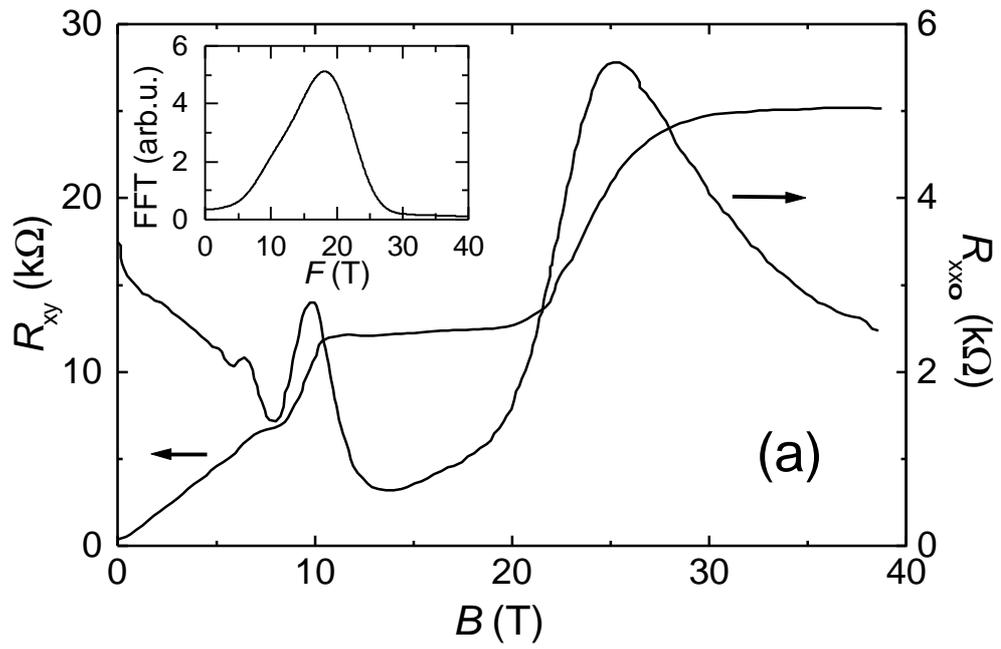

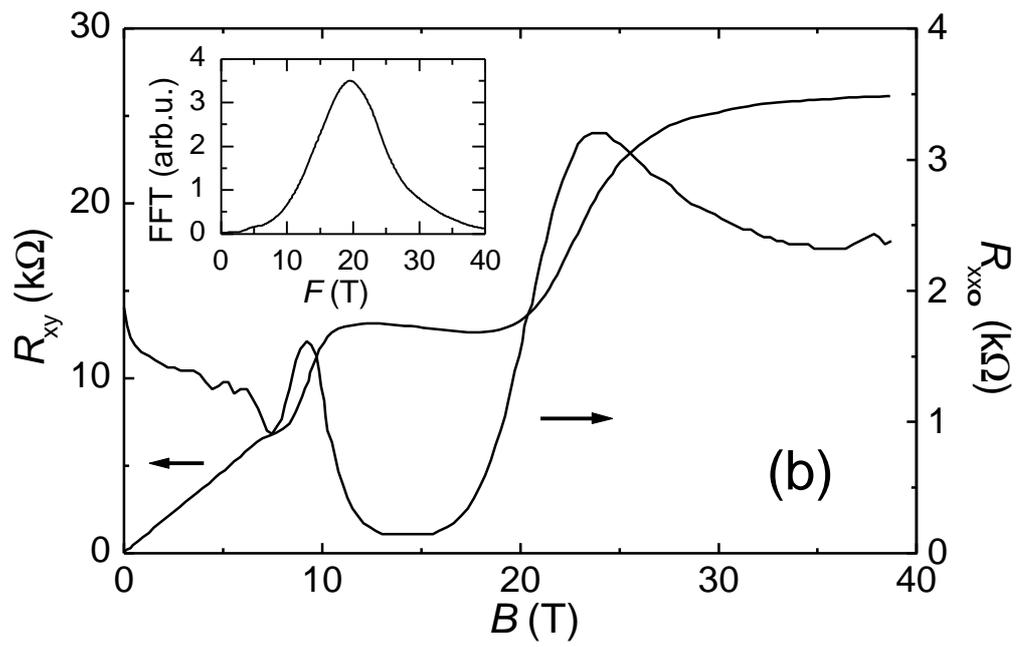

**Fig. 7**